\documentclass{article}
\usepackage{eurosym}
\usepackage{amssymb}
\usepackage{amsmath}
\usepackage{cite}
\usepackage{graphicx}
\usepackage{epsf}

\newtheorem{theorem}{Theorem}

\newtheorem{proposition}[theorem]{Proposition}

\input{tcilatex}

\begin{document}

\title{Lie symmetry analysis and similarity solutions for the Camassa-Choi
equations}
\author{Andronikos Paliathanasis\thanks{%
Email: anpaliat@phys.uoa.gr} \\
{\ \textit{Institute of Systems Science, Durban University of Technology }}\\
{\ \textit{PO Box 1334, Durban 4000, Republic of South Africa}}}
\maketitle

\begin{abstract}
The method of Lie symmetry analysis of differential equations is applied to
determine exact solutions for the Camassa-Choi equation and its
generalization. We prove that the Camassa-Choi equation is invariant under
an infinite-dimensional Lie algebra, with an essential five-dimensional Lie
algebra. The application of the Lie point symmetries leads to the
construction of exact similarity solutions.

\end{abstract}

\section{Introduction}

The Camassa-Choi (CC) equation 
\begin{equation}
\left( u_{t}+\alpha u_{x}-uu_{x}+u_{xx}\right) _{x}+u_{yy}=0.  \label{cc.01}
\end{equation}%
was derived by Choi and Camassa in \cite{choi}$~$in order to describe weakly
nonlinear internal waves in a two-fluid system. Parameter $\alpha =h^{-1}$
describes the depth in the two-fluid system. CC equation can be seen as the
two-dimensional extension of the Benjamin-Ono equation, indeed when $u_{yy}=0
$, from (\ref{cc.01}) the Benjamin-Ono equation is recovered. Because of the
nonlinearity of (\ref{cc.01}) there are not any known-exact solutions in the
literature. Only recently the existence of small data global solutions was
proven by Harrop and Marzula in \cite{cho2}.

In this work, we propose that we apply the theory of Lie point symmetries in
order to determine similarity solutions for the CC equation. The theory of
Lie symmetries of differential equations is the standard technique for the
computation of solutions and describes the algebra for nonlinear
differential equations. The novelty of the Lie symmetries is that invariant
transformations can be found in order to simplify the given differential
equation \cite{app4,ap11,ap12,app5,app6,app7,app8}. There is a plethora of
applications in the Lie symmetries in the fluid dynamics with important
results which have been used to understand the physical properties of the
models.

The Lie symmetry analysis of the Camassa-Holm equation has been previously
performed in \cite{c01}. Similarity solutions to the shallow water equations
with a variable bottom were found in \cite{c02}, while the Lie symmetry
analysis of the rotating shallow-water equation was performed in \cite{c03}.
The algebraic properties of the Benjamin-Ono equation were studied in \cite%
{bo.01}. For other applications in the point symmetries in fluid dynamics we
refer the reader to \cite{c04,c05,c06,c08} and references therein.

We extend our analysis to the generalized Camassa-Choi (GCC) equation%
\begin{equation}
\left( u_{t}+\alpha u_{x}-u^{n}u_{x}+\beta u_{xx}\right) _{x}+u_{yy}=0.
\label{cc.02}
\end{equation}%
which is the natural generalization of the generalized Benjamin-Ono equation 
\cite{b0.3}. For the two equations (\ref{cc.01}), (\ref{cc.02}) we determine
the Lie point symmetries while we prove the existence of travel-wave
similarity solutions for every value of parameter $n\geq 1$ and arbitrary
depth $\alpha $. The plan of the paper is as follows.

In Section \ref{sec2}, for the convenience \ of the reader we briefly
discuss the basic properties and definitions of the theory of Lie point
symmetries. Sections \ref{sec3} and \ref{sec4}, include the main new
material of our analysis, where we present the algebraic properties for
equations (\ref{cc.01}) and (\ref{cc.02}). Finally in Section \ref{sec5} we
draw our conclusions.

\section{Preliminaries}

\label{sec2}

In this section, we briefly discuss the theory of Lie point symmetries of
differential equations which is the main mathematical tool that we apply in
the following.

Consider function $\Phi $ to describe the map of a one-parameter point
transformation such as $u^{\prime }\left( t^{\prime },x^{\prime },y^{\prime
}\right) =\Phi \left( u\left( t,x,y\right) ;\varepsilon \right) \ $where the
infinitesimal transformation is expressed as follows%
\begin{eqnarray}
t^{\prime } &=&t+\varepsilon \xi ^{t}\left( t,x,y,u\right)  \label{sv.12} \\
x^{\prime } &=&x+\varepsilon \xi ^{x}\left( t,x,y,u\right) \\
y^{\prime } &=&y+\varepsilon \xi ^{y}\left( t,x,y,u\right) \\
u &=&u+\varepsilon \eta \left( t,x,y,u\right)  \label{sv.13}
\end{eqnarray}%
where $\varepsilon $ is the infinitesimal parameter, that is, $\varepsilon
^{2}\rightarrow 0.$

From the latter one-parameter point transformation we can define the
infinitesimal generator 
\begin{equation}
X=\frac{\partial t^{\prime }}{\partial \varepsilon }\partial _{t}+\frac{%
\partial x^{\prime }}{\partial \varepsilon }\partial _{x}+\frac{\partial
y^{\prime }}{\partial \varepsilon }\partial _{y}+\frac{\partial u^{\prime }}{%
\partial \varepsilon }\partial _{u},  \label{sv.16}
\end{equation}%
from where the map $\Phi $ can be written as follows 
\begin{equation}
\Phi \left( u\left( t,x,y\right) ;\varepsilon \right) =u\left( t,x,y\right)
+\varepsilon X\left( u\left( t,x,y\right) \right) ,
\end{equation}%
that is,\qquad 
\begin{equation}
X\left( u\left( t,x,y\right) \right) =\lim_{\varepsilon \rightarrow 0}\frac{%
\Phi \left( u\left( t,x,y\right) ;\varepsilon \right) -u\left( t,x,y\right) 
}{\varepsilon }.
\end{equation}

The latter expression defines the Lie derivative of the function $u\left(
t,x,y\right) $ with respect to the vector field $X,$ also noted as $L_{X}u.$

When 
\begin{equation}
L_{X}u=0
\end{equation}%
then we shall say that $u\left( t,x,y\right) $ is invariant under the action
of the one-parameter point transformation with generator the vector field $%
X. $

In terms of differential equations, i.e. 
\begin{equation}
\mathcal{H}\left( u,u_{t},u_{x},u_{y},..\right) =0;
\end{equation}%
then the symmetry condition reads%
\begin{equation}
L_{X}\left( \mathcal{H}\right) =0\text{ or }X^{\left[ n\right] }\left( 
\mathcal{H}\right) =0,  \label{ss1}
\end{equation}%
where $X^{\left[ n\right] }$ describes the $n$th prolongation/extension of
the symmetry vector in the jet-space of variables $\left\{
u,u_{t},u_{x},...\right\} $ defined as%
\begin{equation*}
X^{\left[ n\right] }=X+\eta _{i}^{\left[ 1\right] }\partial
_{u_{i}}+...+\eta ^{\left[ n\right] }\partial _{u_{i_{i}i_{j}...i_{n}}},
\end{equation*}%
where $u_{i}=\frac{\partial u}{\partial z^{i}},~z^{i}=\left( t,x,y\right) $
and%
\begin{equation}
\eta _{i}^{\left[ n\right] }=D_{i}\eta ^{\left[ n-1\right]
}-u_{ii_{2}...i_{n-1}}D_{j}\left( \xi ^{j}\right) ~,~i\succeq 1.
\end{equation}

The main application of the Lie point symmetries is based on the
determination of the Lie invariants which are used to define similarity
transformations and simplify the given differential equation. The exact
solutions which follow by the application of the Lie point symmetries are
called similarity solutions.

If $X$ is an admitted Lie point symmetry, the solution of the associated
Lagrange's system,%
\begin{equation}
\frac{dt}{\xi ^{t}}=\frac{dx}{\xi ^{x}}=\frac{dy}{\xi ^{y}}=\frac{du}{\eta },
\end{equation}%
provides the zeroth-order invariants,~$U^{A\left[ 0\right] }\left(
t,x,u\right) $ which are applied to reduce the number of independent
variables in partial differential equations, or the order in the case of
ordinary differential equations.

For more details on the symmetry analysis of differential equations we refer
the reader to the standard references \cite{book1,book2,book3}.

\section{Point symmetries of the Camassa-Choi equation}

\label{sec3}

From the symmetry condition (\ref{ss1}) for the CC equation (\ref{cc.01})
and for the one-parameter point transformation with generator $X=\xi
^{1}\left( t,x,y,u\right) \partial _{t}+\xi ^{2}\left( t,x,y,u\right)
\partial _{x}+\xi ^{3}\left( t,x,y,u\right) \partial _{y}+\eta \left(
t,x,y,u\right) \partial _{u},$ we find the following system of differential
equations%
\begin{equation}
\xi _{,u}^{1}=0~,~\xi _{,u}^{2}=0~,~\xi _{,u}^{3}=0~,~\xi _{,x}^{1}=0~,~\xi
_{,x}^{3}=0~,~\eta _{,uu}=0~,
\end{equation}%
\begin{equation}
\xi _{,x}^{3}+2\xi _{y}^{1}=0~,~\xi _{,y}^{3}+3\xi _{,x}^{2}=0~,~2\xi
_{,x}^{2}-\xi _{,t}^{1}=0~,~~2\xi _{,y}^{2}-\xi _{,t}^{3}=0,
\end{equation}%
\begin{equation}
\eta _{,xxx}+\left( \alpha -u\right) \eta _{,xx}+\eta _{,yy}+\eta _{,tx}=0~,
\end{equation}%
\begin{equation}
\eta _{,xu}-\xi _{,yy}^{1}=0~,~2\eta _{,yu}-\xi _{,yy}^{3}=0~,~
\end{equation}%
\begin{equation}
3\eta _{,xu}-3\xi _{,xx}^{2}+\left( \alpha -u\right) \xi _{,x}^{2}-\xi
_{,t}^{2}-\eta =0~,
\end{equation}%
\begin{equation}
3\eta _{,xuu}+\left( \alpha -u\right) \eta _{,uu}-\xi _{,x}^{2}-\eta
_{,u}=0~,
\end{equation}%
\begin{equation}
3\eta _{,xxu}-\xi _{,xxx}^{2}+\left( \alpha -u\right) \left( 2\eta
_{,xu}-\xi _{,xx}^{2}\right) +\eta _{,tu}-\xi _{,yy}^{2}-2\eta _{,x}-\xi
_{,tx}^{2}=0.
\end{equation}%
The generic solution of the latter system is $X=\left( c_{1}+c_{2}\left(
2t\right) \right) \partial _{t}+\left( c_{2}x+c_{3}\phi \left( t\right) -%
\frac{1}{2}c_{4}\psi _{t}\left( t\right) y\right) \partial _{x}+\left( \frac{%
3}{2}c_{2}y+c_{4}\psi \left( t\right) \right) \partial _{y}+\left(
c_{2}\left( \alpha -u\right) -c_{3}\phi _{t}\left( t\right) +c_{4}\frac{1}{2}%
\psi _{tt}\left( t\right) y\right) \partial _{u}\,,$ where $%
c_{1},c_{2},c_{3},c_{4}$ are constants of integration and $\phi \left(
t\right) ,~\psi \left( t\right) $ are arbitrary functions.

Therefore, the Lie point symmetries of the CC equation (\ref{cc.01}) are 
\begin{equation}
X_{1}=\partial _{t}~,~X_{2}=2t\partial _{t}+x\partial _{x}+\frac{3}{2}%
y\partial _{y}-\left( u-\alpha \right) \partial _{u}~,  \label{cc.03}
\end{equation}%
\begin{equation}
X_{3}\left( \phi \right) =\phi \left( t\right) \partial _{x}-\phi _{t}\left(
t\right) \partial _{u}~,~X_{4}\left( \psi \right) =\psi \left( t\right)
\partial _{y}-\frac{1}{2}\psi _{t}\left( t\right) y\partial _{x}+\frac{1}{2}%
\psi _{tt}\left( t\right) y\partial _{u}~.  \label{cc.04}
\end{equation}%
Surprisingly, the CC equation admits infinity Lie point symmetries.

The commutators of the Lie point symmetries are 
\begin{eqnarray}
\left[ X_{1},X_{2}\right] &=&2X_{1}~,~\left[ X_{1},X_{3}\left( \phi \right) %
\right] =\left( X_{3}\left( \phi _{t}\right) \right) ~,~\left[
X_{1},X_{4}\left( \psi \right) \right] =\left( X_{4}\left( \psi _{t}\right)
\right) ~,~  \label{cc.05} \\
\left[ X_{2},X_{3}\left( \phi \right) \right] &=&X_{3}\left( \phi -2t\phi
_{t}\right) ~,~\left[ X_{2},X_{4}\left( \psi \right) \right] =X_{4}\left( 
\frac{3}{4}\psi -2t\psi _{t}\right) ~,~\left[ X_{3}\left( \phi \right)
,X_{4}\left( \psi \right) \right] =0~,  \label{cc.06} \\
\left[ X_{3}\left( \phi \right) ,X_{3}^{\prime }\left( \chi \right) \right]
&=&0~\ ,~\left[ X_{4}\left( \psi \right) ,X_{4}\left( \xi \right) \right] =%
\frac{1}{2}\left( X_{3}\left( \xi \psi _{t}-\psi \xi _{t}\right) \right) .
\label{cc.07}
\end{eqnarray}%
from where we observe that they form an infinity-dimensional Lie algebra.
The existence of the infinity number of symmetries it is not a real
surprise. From $X_{3}$ we determine the similarity transformation $u=$ $%
-\left( \ln \phi \left( t\right) \right) _{,t}x+U\left( t,y\right) $ where $%
U\left( t,y\right) =\frac{1}{2}\frac{\phi _{,tt}}{\phi }y^{2}+U_{1}\left(
t\right) y+U_{0}\left( t\right) $ solves the reduced equation, functions $%
U_{1}\left( t\right) ,~U_{0}\left( t\right) $ are arbitrary functions.

In the special case where $\phi \left( t\right) $ and $\psi \left( t\right) $
are constants, without loss of generality we assume that $\phi \left(
t\right) =\psi \left( t\right) =1$, the Lie point symmetries are simplified
as 
\begin{equation}
X_{1}^{\prime }=\partial _{t}~,~X_{2}^{\prime }=2t\partial _{t}+x\partial
_{x}+\frac{3}{2}y\partial _{y}-\left( u-\alpha \right) \partial
_{u}~,~X_{3}^{\prime }=\partial _{x}~,~X_{4}^{\prime }=\partial _{y}
\label{cc.08}
\end{equation}%
with commutators%
\begin{eqnarray}
\left[ X_{1}^{\prime },X_{2}^{\prime }\right] &=&2X_{1}^{\prime }~,\left[
X_{1}^{\prime },X_{3}^{\prime }\right] =0~,~\left[ X_{1}^{\prime
},X_{4}^{\prime }\right] =0~,~  \label{cc.09} \\
\left[ X_{2}^{\prime },X_{3}^{\prime }\right] &=&X_{3}^{\prime }~,~\left[
X_{2}^{\prime },X_{4}^{\prime }\right] =\frac{3}{2}X_{3}^{\prime }~,~\left[
X_{3}^{\prime },X_{4}^{\prime }\right] =0.  \label{cc.10}
\end{eqnarray}%
However, there are not any finite-dimensional closed Lie algebras for
arbitrary functions of $\phi \left( t\right) $ and $\psi \left( t\right) $.
The commutators of the latter finite-dimensional Lie algebra are presented
in Table \ref{tab1}.

\begin{table}[tbp] \centering%
\caption{Commutators for Lie point symmetries of CC which form a
finite-dimensional Lie algebra}%
\begin{tabular}{c|cccc}
\hline\hline
$\left[ ~,~\right] $ & $X_{1}^{\prime }$ & $X_{2}^{\prime }$ & $%
X_{3}^{\prime }$ & $X_{4}^{\prime }$ \\ \hline
$X_{1}^{\prime }$ & $0$ & $2X_{1}^{\prime }$ & $0$ & $\,0$ \\ 
$X_{2}^{\prime }$ & $2X_{1}^{\prime }$ & $0$ & $X_{3}^{\prime }$ & $\frac{3}{%
2}X_{3}^{\prime }$ \\ 
$X_{3}^{\prime }$ & $0$ & $-X_{3}^{\prime }$ & $0$ & $0$ \\ 
$X_{4}^{\prime }$ & $0$ & $-\frac{3}{2}X_{3}^{\prime }$ & $0$ & $0$ \\ 
\hline\hline
\end{tabular}%
\label{tab1}%
\end{table}%

Let us demonstrate that by assuming $\phi \left( t\right) =\phi _{1}+\phi
_{2}e^{\omega _{1}t}~$and $\psi \left( t\right) =\psi _{1}+\psi
_{2}e^{\omega _{2}t}.$ Then from (\ref{cc.03}), (\ref{cc.04}) it follows
that the CC equation admits six Lie point symmetries which are the vector
fields 
\begin{equation}
X_{1}^{\prime }~,~X_{2}^{\prime }~,~X_{3}^{\prime }~,~X_{4}^{\prime
}~,~X_{5}^{\prime }=e^{\omega _{1}t}\left( \partial _{x}-\omega _{1}\partial
_{u}\right) ~,~X_{6}^{^{\prime }}=e^{\omega _{1}t}\left( \partial _{y}-\frac{%
\omega _{2}}{2}y\partial _{x}+\frac{\omega _{2}^{2}}{2}y\partial _{u}\right)
\label{cc.11}
\end{equation}%
with commutators (\ref{cc.09}), (\ref{cc.10}) and 
\begin{eqnarray}
\left[ X_{1}^{\prime },X_{5}^{\prime }\right] &=&\omega _{1}X_{5}^{\prime
}~,~\left[ X_{1}^{\prime },X_{6}^{\prime }\right] =\omega _{2}X_{6}^{\prime
}~,~\left[ X_{2}^{\prime },X_{5}^{\prime }\right] =e^{\omega _{1}t}\left(
\left( 1-\omega _{1}t\right) \partial _{x}+\omega _{1}\left( 1+2\omega
_{1}t\right) \partial _{u}\right) ~,  \label{cc.12} \\
\left[ X_{2}^{\prime },X_{6}^{\prime }\right] &=&e^{\omega _{2t}}\left( 
\frac{\left( 3-4\omega _{2}t\right) }{2}\partial _{y}+\frac{\left( 1+4\omega
_{2}t\right) }{4}\omega _{2}y\partial _{x}-\frac{\left( 5+4\omega
_{2}t\right) }{4}\omega _{2}^{2}y\partial _{u}\right) ~,  \label{cc.13} \\
~\left[ X_{3}^{\prime },X_{5}^{\prime }\right] &=&0~,~\left[ X_{3}^{^{\prime
}},X_{6}^{^{\prime }}\right] =0~,~\left[ X_{4}^{\prime },X_{6}^{\prime }%
\right] =\frac{\omega _{2}}{2}e^{\omega _{2}t}\left( \partial _{t}-\omega
_{2}\partial _{u}\right) .  \label{cc.14}
\end{eqnarray}%
from where it is clear that the symmetry vectors (\ref{cc.11}) do not form a
closed Lie algebra.

We want to constraint functions $\phi \left( t\right) ,~$and $\psi \left(
t\right) $ such that the admitted Lie symmetries to form a closed Lie
algebra of five-dimension with a different basis. In particular we focus on
the case where the coefficients of the commutators (\ref{cc.05})-(\ref{cc.07}%
) are constants. \ Thus we end up with the system of equations 
\begin{equation}
\left\{ \phi =c_{1}\phi _{t}~,~\phi =c_{2}\phi -2t\phi _{t}\right\} ~~\text{%
or~}\left\{ \phi =c_{1}^{\prime }\phi _{t}~,~\phi _{t}=c_{2}^{\prime }\left(
\phi -2t\phi _{t}\right) \right\} ~,  \label{cc.15}
\end{equation}%
and%
\begin{equation}
\left\{ \psi =c_{3}\psi _{t}~,~\psi =c_{4}\left( \frac{3}{4}\psi -2t\psi
_{t}\right) \right\} \text{ or~}\left\{ \psi ^{\prime }=c_{3}\psi
_{t}~,~\psi _{t}^{\prime }=c_{4}\left( \frac{3}{4}\psi -2t\psi _{t}\right)
\right\} ,  \label{cc.16}
\end{equation}%
with constraint equations%
\begin{equation}
\xi =\xi \psi _{t}-\psi \xi _{t}\text{, where }\xi =\phi ,~\text{or ~}\xi
=\phi _{t}\text{ or }\xi =\left( \phi -2t\phi _{t}\right) .  \label{cc.17}
\end{equation}

Therefore, from (\ref{cc.15}), (\ref{cc.16}) and (\ref{cc.17}) it follows
that the unique possible admitted five-dimensional Lie algebra is that of (%
\ref{cc.08}) for $\phi \left( t\right) =const$. $\ $and $\psi \left(
t\right) =\psi _{0}+\psi _{1}t$. Of course there are additional finite
dimensional Lie algebras, for instance any set of generators$~$constructed
by $X_{3}$ form a Lie algebra; however this specific five-dimensional Lie
algebra has the novelty that\ it can provide a plethora of different
similarity transformations, while for instance the similarity
transformations which follow by $X_{3}$ are all of the same family.

\begin{proposition}
The CC equation (\ref{cc.01}) is invariant under infinity Lie point
symmetries which form the Lie algebra $\left\{ A_{2,1}\otimes _{s}A_{\infty
}\otimes _{s}A_{\infty }\right\} $ in the Morozov-Mubarakzyanov
classification scheme \cite{mb1,mb2,mb3,mb4}. However, there exists a
five-dimensional subalgebra consisted by the vector fields $\left\{
X_{1},X_{2},X_{3}^{\prime },X_{4}^{\prime },X_{5}=t\partial _{y}-\frac{1}{2}%
y\partial _{x}\right\} $ and form the Lie algebra $A_{5,19}^{ab}$ in the
Patera-Winternitz classification scheme \cite{pat11,pat12}. This
five-dimensional Lie algebra provides the maximum number of alternative
families of similarity transformations.
\end{proposition}

As we shall see in the following, this five-dimensional Lie algebra plays a
significant role in the study of the Lie point symmetries for the GCC
equation (\ref{cc.02}). We proceed with the application of the Lie point
symmetries for the derivation of similarity solutions.

\subsection{Similarity Solutions for the Camassa-Choi equation}

Let us not apply the Lie point symmetries found in the previous section in
order to find similarity solutions for the CC equation (\ref{cc.01}). The CC
equation is a third equation of three independent variables. By applying the
Lie point symmetries in partial differential equations we reduce the number
of the independent variables. Hence, in order to reduce the CC equation to
an ordinary differential equation we should apply two symmetry vectors.
However, not all the symmetry vectors survive through the reduction process.
In particular, if a given differential equation admits the two symmetry
vectors $\Gamma _{1},\Gamma _{2}$ with commutator $\left[ \Gamma _{1},\Gamma
_{2}\right] =c\Gamma _{2}$, where $c$ may be zero, then reduction of the
differential equation with respect to the symmetry vector $\Gamma _{2}$
provides that the reduced equation inherits the symmetry vector $\Gamma _{1}$%
, while reduction with $\Gamma _{1}$ provides a differential equation where $%
\Gamma _{2}$ is not a point symmetry when $c\neq 0~$\cite{Govinder2001}. It
is clear, that if we want to perform a second reduction for the differential
equation we start by considering the symmetry vector $\Gamma _{2}$.

Therefore, by using the results of Table \ref{tab1} we find that the
reduction with the symmetry vectors $\left\{ X_{1}^{\prime },X_{3}^{\prime
},X_{4}^{\prime },X_{3}^{\prime }+X_{4}^{\prime }\right\} $ gives reduced
equations which inherits some of the symmetries of the original equation.
However, the application of the symmetry vectors $\left\{ X_{1}^{\prime
},X_{3}^{\prime },X_{4}^{\prime }\right\} $ gives time-independent or static
solutions which are not solutions of special interests solutions. Hence, we
focus on the reduction which follows by the symmetry vector $X_{3}^{\prime
}+X_{4}^{\prime }$.

From the Lie point symmetry $X_{34}=X_{3}^{\prime }+X_{4}^{\prime }$ we
calculate the invariants 
\begin{equation}
t~,~w=y-x~,~u=U\left( t,w\right) .  \label{cc.18}
\end{equation}%
By using the latter invariant functions equation (\ref{cc.01}) is reduced to
the following partial differential equation%
\begin{equation}
U_{www}+\left( U_{w}\right) ^{2}-\left( 1-U+h_{0}\right) U_{ww}+U_{wt}=0.
\label{cc.19}
\end{equation}

In order to proceed with the reduction we should derive the Lie point
symmetries of (\ref{cc.19}). Hence, by applying the Lie symmetry condition
we find that equation (\ref{cc.19}) is invariant under the Lie point
symmetries%
\begin{eqnarray}
Z_{1} &=&\partial _{t},~Z_{2}=2t\partial _{t}+w\partial _{w}+\left(
h_{0}+1-U\right) \partial _{U}~,  \label{cc.20} \\
Z_{3} &=&t^{2}\partial _{t}+tw\partial _{w}+\left[ \left( h_{0}+1-U\right)
t+w\right] \partial _{U}~,  \label{cc.21} \\
Z_{4} &=&\phi \left( t\right) \partial _{w}+\phi _{t}\partial _{U}.
\label{cc.23}
\end{eqnarray}%
Vector fields $Z_{1},~Z_{2}$ and $Z_{4}$ are reduced symmetries, while $%
Z_{3} $ is a new symmetry for the reduced equation (\ref{cc.19}). It is
important to mention that $Z_{4}$ describes an infinity number of
symmetries, hence the reduced equation (\ref{cc.19}) admits infinity number
of Lie point symmetries as the \textquotedblleft mother\textquotedblright\
equation (\ref{cc.01}). On the other hand, Lie point symmetries $\left\{
Z_{1},Z_{2},Z_{3}\right\} $ form a closed Lie algebra, known as $SL\left(
2,R\right) $.

The application of $Z_{4}$ in (\ref{cc.19}) provides the linear second-order
ODE $\phi _{tt}=0$, where $U\left( t,w\right) =U_{0}\left( t\right) +\frac{%
\phi _{t}}{\phi }w,$ where $U_{0}\left( t\right) $ is an arbitrary function.
Therefore, the similarity solution is derived to be 
\begin{equation}
U\left( t,w\right) =U_{0}\left( t\right) +\frac{\phi _{1}}{\phi _{1}t+\phi
_{0}}w.  \label{cc.24}
\end{equation}

Reduction with respect the symmetry vector $Z_{1}$ of equation (\ref{cc.19})
provides the third-order ODE%
\begin{equation}
Y_{www}+\left( Y_{w}\right) ^{2}-YY_{ww}=0~,~U\left( t,w\right) =Y\left(
w\right) +1+h_{0}\,~,w=x  \label{cc.25}
\end{equation}%
which admit two point symmetries the reduced symmetries $Z_{2},~$and $Z_{3}$%
. Equation (\ref{cc.25}) can be integrated as follows%
\begin{equation}
Y_{ww}+YY_{w}+Y_{0}=0,  \label{cc.26}
\end{equation}%
where the latter equation can be solved in terms of quadratics. Indeed for
the integration constant $Y_{0}=0$, the general solution is 
\begin{equation}
Y\left( w\right) =Y_{0}\tanh \left( \frac{w-w_{0}}{2c}\right) ,
\label{cc.27}
\end{equation}%
while in general equation (\ref{cc.26}) becomes 
\begin{equation}
Y_{w}+\frac{1}{2}Y^{2}+Y_{0}w+Y_{1}=0.  \label{cc.28}
\end{equation}

The application of the Lie symmetry vector $Z_{2}$ provides the reduced
third-order ODE%
\begin{equation}
2\bar{Y}_{\sigma \sigma \sigma }+\left( \sigma -2\left( 1+\bar{Y}\right)
\right) \bar{Y}_{\sigma \sigma }-2\bar{Y}=0~,~  \label{cc.29}
\end{equation}%
where now $U\left( t,w\right) =1+h_{0}+\frac{Y\left( \sigma \right) }{\sqrt{t%
}}~,~\sigma =\frac{w}{\sqrt{t}}$. The latter equation can be easily
integrated as follows%
\begin{equation}
2\bar{Y}_{\sigma \sigma }-\bar{Y}-\left( 2\bar{Y}-\sigma \right) \bar{Y}+%
\bar{Y}_{0}=0  \label{cc.30}
\end{equation}%
or%
\begin{equation}
2\bar{Y}_{\sigma }+\bar{Y}^{2}-\sigma \bar{Y}+\bar{Y}_{0}\sigma +\bar{Y}%
_{1}=0.  \label{cc.31}
\end{equation}

In a similar way, the reduction with respect to the Lie symmetry vector $%
Z_{3}$ gives the solution%
\begin{equation}
U\left( t,w\right) =\frac{w}{t}+h_{0}+1+\frac{Y^{\prime }\left( \lambda
\right) }{t}~,~\lambda =\frac{w}{t},  \label{cc.32}
\end{equation}%
where $Y^{\prime }\left( \lambda \right) $ is given by the following
first-order ODE%
\begin{equation}
Y_{\lambda }^{\prime }+\frac{1}{2}\left( Y^{\prime }\right)
^{2}+Y_{0}\lambda +Y_{1}=0.  \label{cc.33}
\end{equation}

It comes as no surprise that the reduction with the three elements of the $%
SL\left( 2,R\right) $ provides similar reduced equations. That is because
the three symmetry vectors are related with similarity transformations as
well as also the reduced equations are related, for more details we refer
the reader to \cite{leach1}.

Finally, reduction with the vector field $Z_{1}+Z_{4}$, for $\phi \left(
t\right) =1$, provides travel-wave solution and the reduced equation is that
of (\ref{cc.25}) \ where $w=t-x$.

Similarly, the reduction of CC equation (\ref{cc.01}) with respect the
symmetry vector $X_{14}=X_{1}^{\prime }+X_{4}^{\prime }$, provides a
travel-wave solution, as before. Therefore, we conclude that travel-wave
solutions exist for the CC equation.

We proceed our analysis by studying the invariant point transformations for
the GCC equation (\ref{cc.02}).

\section{Point symmetries of the generalized Camassa-Choi equation}

\label{sec4}

The Lie point symmetries of the GCC equation (\ref{cc.02}) are%
\begin{equation}
Y_{1}=\partial _{t}~,~Y_{2}=2t\partial _{t}+x\partial _{x}+\frac{3}{2}%
y\partial _{y}-\frac{1}{n}u\partial _{u}
\end{equation}%
\begin{equation}
Y_{3}=\partial _{x}~,~Y_{4}=\partial _{y}~,~Y_{5}=2t\partial _{y}-y\partial
_{x}~,
\end{equation}%
when $\alpha =0$ and 
\begin{equation}
\bar{Y}_{1}=\partial _{t}~,~\bar{Y}_{2}=\partial _{x}~,\bar{Y}%
_{2}=2t\partial _{t}+\left( x+\alpha t\right) \partial _{x}+\frac{3}{2}%
y\partial _{y}-\frac{1}{n}u\partial _{u}
\end{equation}%
\begin{equation}
\bar{Y}_{3}=\partial _{x}~,~\bar{Y}_{4}=\partial _{y}~,~\bar{Y}%
_{5}=2t\partial _{y}-y\partial _{x}~,
\end{equation}%
for $\alpha \neq 0$. \ 

The corresponding commutators for the admitted Lie symmetries are presented
in Table \ref{tab2}. We observe that the two admitted Lie algebras are
different. For $\alpha \neq 0$ the Lie symmetries form the Lie algebra $%
A_{5,23}^{b}$ and for $\alpha =0$, the Lie symmetries form the Lie algebra $%
A_{5,19}^{ab}$ in the Patera-Winternitz classification scheme \cite%
{pat11,pat12}.

\begin{table}[tbp] \centering%
\caption{Commutators of the admitted Lie point symmetries by the GCC}%
\begin{tabular}{c|ccccc}
\hline\hline
$\left[ ~,~\right] $ & $\bar{Y}_{1}$ & $\bar{Y}_{2}$ & $\bar{Y}_{3}$ & $\bar{%
Y}_{4}$ & $\bar{Y}_{5}$ \\ \hline
$\bar{Y}_{1}$ & $0$ & $2Y_{1}+\alpha Y_{3}$ & $0$ & $0$ & $2\bar{Y}_{4}$ \\ 
$\bar{Y}_{2}$ & $-\left( 2\bar{Y}_{1}+\alpha Y_{3}\right) $ & $0$ & $-\bar{Y}%
_{3}$ & $-\frac{3}{2}\bar{Y}_{4}$ & $\frac{1}{2}\bar{Y}_{5}$ \\ 
$\bar{Y}_{3}$ & $0$ & $\bar{Y}_{3}$ & $0$ & $0$ & $0$ \\ 
$\bar{Y}_{4}$ & $0$ & $\frac{3}{2}\bar{Y}_{4}$ & $0$ & $0$ & $-\bar{Y}_{3}$
\\ 
$\bar{Y}_{5}$ & $-2\bar{Y}_{4}$ & $-\frac{1}{2}\bar{Y}_{5}$ & $0$ & $\bar{Y}%
_{3}$ & $0$ \\ \hline\hline
\end{tabular}%
\label{tab2}%
\end{table}%

When the parameter $\alpha $ is zero, the Lie point symmetries $\left\{
Y_{1},Y_{2},Y_{3},Y_{4}\right\} $ are these which form a finite-dimensional
Lie algebra for the CC equation (\ref{cc.01}), that is, vector fields (\ref%
{cc.08}). However, When $\alpha \neq 0$ things are different. The fifth
symmetry $Y_{5}$ is a case of $X_{4}\left( \psi \right) $ with $\psi \left(
t\right) =t$. Indeed, the admitted Lie point symmetries by the GCC are those
which form the maximum finite-dimensional Lie algebra for the CC equation.

We continue our analysis by applying the Lie point symmetries to determine
similarity solutions for the GCC equation.

\subsection{Similarity Solutions for the generalized Camassa-Choi equation}

As in the case of the CC we consider the similarity transformation provided
by the vector field $Y_{34}=Y_{3}+Y_{4}$, because it is the similarity
transformation which provides a reduced equation which inherits symmetry
vectors. Hence, we find that the GCC equation (\ref{cc.02}) is reduced to%
\begin{equation}
U_{www}+U_{wt}+nU^{n-1}\left( U_{w}\right) ^{2}+\left( U^{n}+1-\alpha
\right) U_{ww}=0,  \label{eq.33}
\end{equation}%
where $u\left( t,x,y\right) =U\left( t,w\right) $ and $w=x-y.~$We observe
that equation (\ref{eq.33}) reduces into (\ref{cc.19}) when $n=1$.

For $n\neq 1,$ we calculate the Lie point symmetries of (\ref{eq.33}) which
are we found to be 
\begin{equation*}
\bar{Z}_{1}=\partial _{t},~\bar{Z}_{2}=\partial _{w}~\text{and }\bar{Z}%
_{3}=2t\partial _{t}+\left( t\left( 1+\alpha \right) -w\right) \partial _{w}-%
\frac{1}{n}U\partial _{u}.
\end{equation*}

The application of the Lie symmetry vector $\bar{Z}_{12}=\partial
_{t}+\partial _{w}$ in (\ref{eq.33}) provides the travel-wave solution 
\begin{equation}
Y_{\sigma \sigma \sigma }+nY^{n-1}\left( Y_{\sigma }\right) ^{2}+\left(
Y^{n}-\alpha -2\right) Y_{\sigma \sigma }=0~,~U\left( t,w\right) =Y\left(
\sigma \right) ~,~\sigma =w-t.  \label{eq.34}
\end{equation}%
The latter equation can be easily integrated by quadratures as follows%
\begin{equation}
Y_{\sigma }+\frac{1}{n+1}Y^{n+1}+\left( 2+A\right) Y+Y_{1}\sigma +Y_{0}=0.
\label{eq.35}
\end{equation}

On the other hand, the reduction of (\ref{eq.33}) with respect to the
similarity transformation provided by the vector field $\bar{Z}_{3}$
provides 
\begin{equation}
U\left( t,w\right) =H\left( \zeta \right) t^{-\frac{1}{2n}}~\ ,~\zeta =\frac{%
w+t\left( 1+\alpha \right) }{\sqrt{t}}  \label{eq.36}
\end{equation}%
where$~H\left( \zeta \right) $ satisfies the third-order ordinary
differential equation 
\begin{equation}
2nHH_{\zeta \zeta \zeta }+nH\left( 2H^{n}-\zeta \right) H_{\zeta \zeta
}-\left( \left( n+1\right) H-2n^{2}H^{n}H_{z}\right) H_{z}=0.  \label{eq.37}
\end{equation}%
Equation (\ref{eq.37}) can be integrated as follows%
\begin{equation}
H_{\zeta \zeta }-\frac{1}{2n}H+\left( H^{n}-\frac{\zeta }{2}H\right)
H_{\zeta }+H_{1}=0.  \label{eq.38}
\end{equation}

\begin{figure}[tbp]
\includegraphics[width=0.4\textwidth]{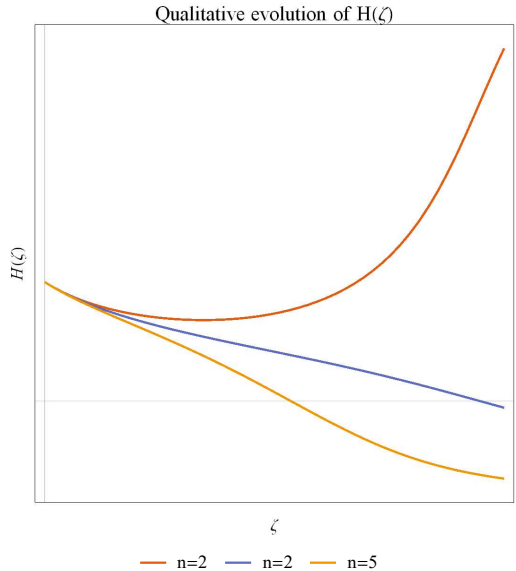}\centering
\caption{Qualitative evolution of$~H(\protect\zeta )$ as it is given by the
differential equation (\protect\ref{eq.38}) for initial conditions $H\left(
0\right) =1$ and $H_{\protect\zeta }\left( 0\right) =-0.5$. The plots are
for $H_{1}=0$ and $n=2~$(red line), $n=3$ (blue line) and $n=5$ (yellow
line). $\ $}
\label{fig0}
\end{figure}

The latter equation does not admit any point symmetry and we cannot perform
further reduction. However in \ref{fig0} we present some numerical
solutions. What is also important to mention is that in equation (\ref{eq.38}%
) parameter $\alpha $ plays no role. Hence the same reduction holds and for
the case $\alpha =0$.

\section{Conclusions}

\label{sec5}

In this work, we applied the theory of symmetries of differential equations
in order to determine exact similarity solutions for the Camassa-Choi
equation (\ref{cc.01}) and its generalization (\ref{cc.02}). CC equation
describes weakly nonlinear internal waves in a two-fluid system and it can
be seen as the two-dimensional generalization of the Benjamin-Ono.

For the CC equation we found that it is invariant under an
infinity-dimensional Lie algebra, with maximum finite Lie subalgebra \ of
dimension five. That five-dimensional subalgebra is the one which form the
complete group of invariant one-parameter point transformations for the GCC
equation.

We apply the Lie point symmetries and we prove the existence of similarity
solutions in the two-dimensional plane $\left\{ x,y\right\} $. Specifically,
we found that the similarity solutions can be expressed in terms of
quadratures.

Surprisingly, the CC equation under the application of similarity
transformations can be reduced into a three-dimensional ordinary
differential equation which is invariant under the $SL\left( 3,R\right) $,
where all the possible reductions provide similarity solutions related under
point transformations.

In a future work we plant to study the physical properties of those new
similarity solutions.

\end{document}